\documentclass[traditabstract]{aa}
\usepackage{txfonts}
\usepackage{graphicx}
\usepackage{natbib}
\bibpunct{(}{)}{;}{a}{}{,} 
\usepackage{epsfig}

\newcommand{\teff}{\mbox{$T_{\rm eff}$}}
\newcommand{\logg}{\mbox{$\log g$}}
\newcommand{\vsini}{\mbox{$v \sin i_\star$}}
\newcommand{\mictrb}{\mbox{$\xi_{\rm t}$}}

\newcommand{\kms}{\mbox{km\,s$^{-1}$}}
\newcommand{\halpha}{\mbox{$H_\alpha$}}

\begin{document}

\title{WASP-80b: a gas giant transiting a cool dwarf\thanks{using WASP-South photometric observations, from Sutherland (South Africa), confirmed with the 60\,cm TRAPPIST robotic telescope, EulerCam, and the CORALIE spectrograph on the Swiss 1.2\,m \textit{Euler} Telescope, and HARPS on the ESO 3.6m (Prog ID 089.C-0151), all three located at La Silla Observatory, Chile. The data is publicly available at the \textit{CDS} Strasbourg and on demand to the main author.}
}

\author{Amaury H.~M.~J. Triaud\inst{1}
\and D.~R. Anderson\inst{2}
\and A.~Collier Cameron\inst{3}
\and A.~P. Doyle\inst{2}
\and A. Fumel\inst{4}
\and M.~Gillon\inst{4} 
\and C.~Hellier\inst{2}
\and E.~Jehin\inst{4}
\and M.~Lendl\inst{1}
\and C.~Lovis\inst{1}
\and P.~F.~L. Maxted\inst{2}
\and F.~Pepe\inst{1}
\and D.~Pollacco\inst{5}
\and D.~Queloz\inst{1}
\and D.~S\'egransan\inst{1}
\and B.~Smalley\inst{2}
\and A.~M.~S. Smith\inst{2,6}
\and S.~Udry \inst{1}
\and R.~G. West\inst{7}
\and P.~J. Wheatley\inst{5}
}

\offprints{Amaury.Triaud@unige.ch}

\institute{Observatoire Astronomique de l'Universit\'e de Gen\`eve, Chemin des Maillettes 51, CH-1290 Sauverny, Switzerland
\and Astrophysics Group, Keele University, Staffordshire, ST55BG, UK
\and SUPA, School of Physics \& Astronomy, University of St Andrews, North Haugh, KY16 9SS, St Andrews, Fife, Scotland, UK
\and Institut d'Astrophysique et de G\'eophysique, Universit\'e de Li\`ege, All\'ee du 6 Ao\^ut, 17, Bat. B5C, Li\`ege 1, Belgium
\and Department of Physics, University of Warwick, Coventry CV4 7AL, UK
\and N. Copernicus Astronomical Centre, Polish Academy of Sciences, Bartycka 18, 00-716 Warsaw, Poland
\and Department of Physics and Astronomy, University of Leicester, Leicester, LE17RH, UK
}

\date{Received date / accepted date}
\authorrunning{Triaud A.\,H.\,M.\,J. et al.}
\titlerunning{WASP-80b}

\abstract{We report the discovery of a planet transiting the star \object{WASP-80} (\object{1SWASP J201240.26-020838.2}; \object{2MASS J20124017-0208391}; \object{TYC~5165-481-1}; \object{BPM 80815}; V=11.9, K=8.4). 
Our analysis shows this is a $0.55\pm0.04$~M$_\mathrm{jup}$, $0.95\pm0.03$~R$_\mathrm{jup}$  gas giant on a circular 3.07 
day orbit  around a star with a spectral type between K7V and M0V. 
This system produces one of the largest transit depths so far reported, making it a worthwhile target for transmission spectroscopy. We find a large 
discrepancy between the \vsini\, inferred from stellar line broadening and the observed amplitude of the Rossiter-McLaughlin effect. This can be understood either by
an orbital plane nearly perpendicular to the stellar spin or by an additional, unaccounted for source of broadening.

\keywords{binaries: eclipsing -- planetary systems -- stars: individual: \object{WASP-80} -- techniques: spectroscopic; photometric} }

\maketitle


\begin{figure*}
\centering
\includegraphics[width=18cm]{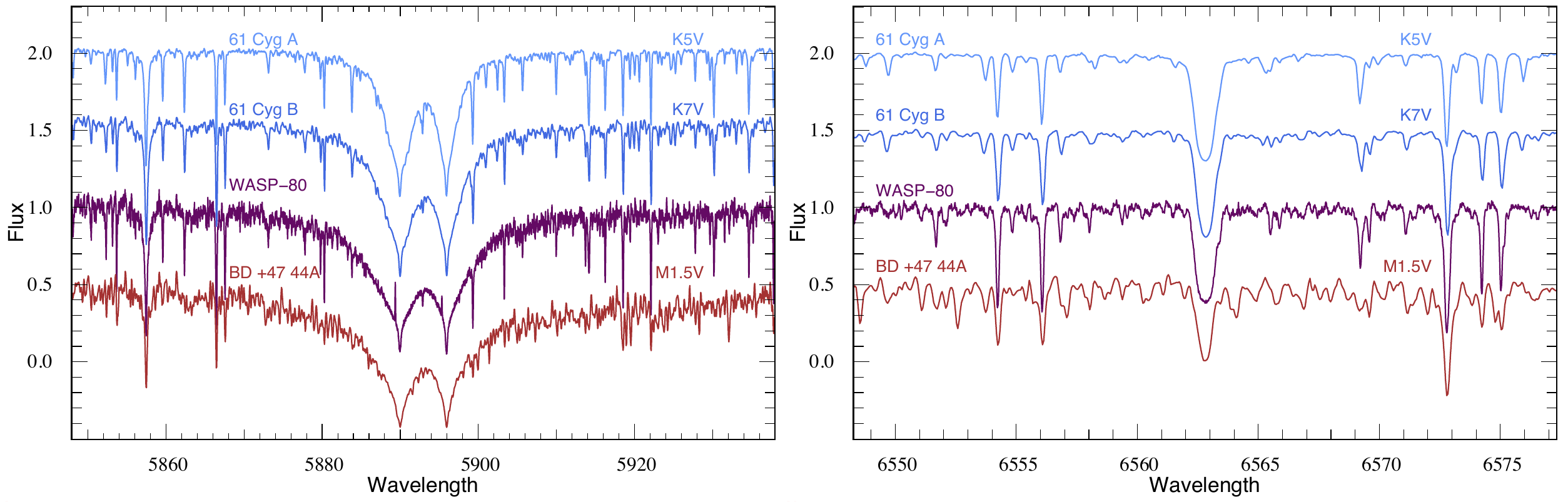}
\caption{Co-added HARPS spectra on \object{WASP-80} compared with ELODIE spectra of 61 Cyg A \& B and of BD\,+47\,44A over the Na {\sc i} Doublet~and~H$_\alpha$.}
\label{fig:spec}
\end{figure*}

Numerous planets have been found since \citet{Mayor:1995uq} and, like 51 Pegasi\,b, most orbit stars whose spectral type, mass, or size 
are similar to the Sun's. This occurs even though a few surveys have concentrated their efforts on other spectral types notably towards M dwarfs, such 
as the HARPS M-dwarf survey \citep{Bonfils:2013fk} or the M-Earth project \citep{Nutzman:2008qy}. In their rarity, those planets nevertheless help us better understand the processes leading to planet formation. 

The WASP (Wide Angle Search for Planets) survey aims to find transiting planets \citep{Pollacco:2006fj}, and has now surveyed most of the night sky in 
both hemispheres. With some 70 planets now publicly announced, this is the most efficient ground-based planet discovery project. Thanks to its observation 
of now more than 30 million stars of magnitude between 8.5 and 13.5, it can pickup those rare planets that have avoided detection by the radial-velocity 
surveys or even by the space-missions \textit{Kepler} and CoRoT which have surveyed \textit{only} 150\,000 stars each. Amongst those rare planets found 
by WASP is the first gas giant around a $\delta$ Scuti \citep{Collier-Cameron:2010lr} and the population of very short period gas giants, such as WASP-12, 
18, 19 and 43 \citep{Hebb:2009lr,Hebb:2010fk,Hellier:2009lr,Hellier:2011fk}.

Despite their numbers and the facility of discovering them (radial velocities or transits), the occurrence rate of hot Jupiters orbiting solar-type stars is  low. 
It has been estimated to be as high as $1.5\pm0.6\%$ by \citet{Cumming:2008ys} from radial velocity surveys, and as low as 
$0.5\pm0.1\%$ by \citet{Howard:2012ul} from the \textit{Kepler} results. \citet{Johnson:2010ly} have made a case that,
 because no hot Jupiter was known to orbit an M dwarf, their occurrence must therefore be lower. Not long 
afterwards \citet{Johnson:2012vn}  announced the discovery of a transiting gas giant around a star observed by \textit{Kepler}, KOI-254, describing it as a 
\textit{"lone example [...] for some time to come"}. Approximately 300 M dwarf systems have been searched for planets between the main radial velocity 
teams \citep{Johnson:2010ly}. The M-Earth project is targeting about 3\,000 (with a geometrical detection of only 5-10\%). If the rate of hot Jupiters is 
but a half to a third that of solar type stars, there is a significant chance that such planets have avoided detection, a point made by \citet{Bonfils:2013fk}. Knowing this rate 
is important since gas giant formation is perceived as less efficient because protoplanetary disc masses scale with their primary's mass  
as  dynamical timescales do \citep{Laughlin:2004gf,Ida:2005lr,Alibert:2011fk,Mordasini:2012ve}.


Within this context, we announce the discovery of a gas giant transiting a late K--early M dwarf. We first describe our data collection, then its analysis, 
and finally the results we obtain.

\section{Observations}

\begin{figure*}
\centering
\includegraphics[width=18cm]{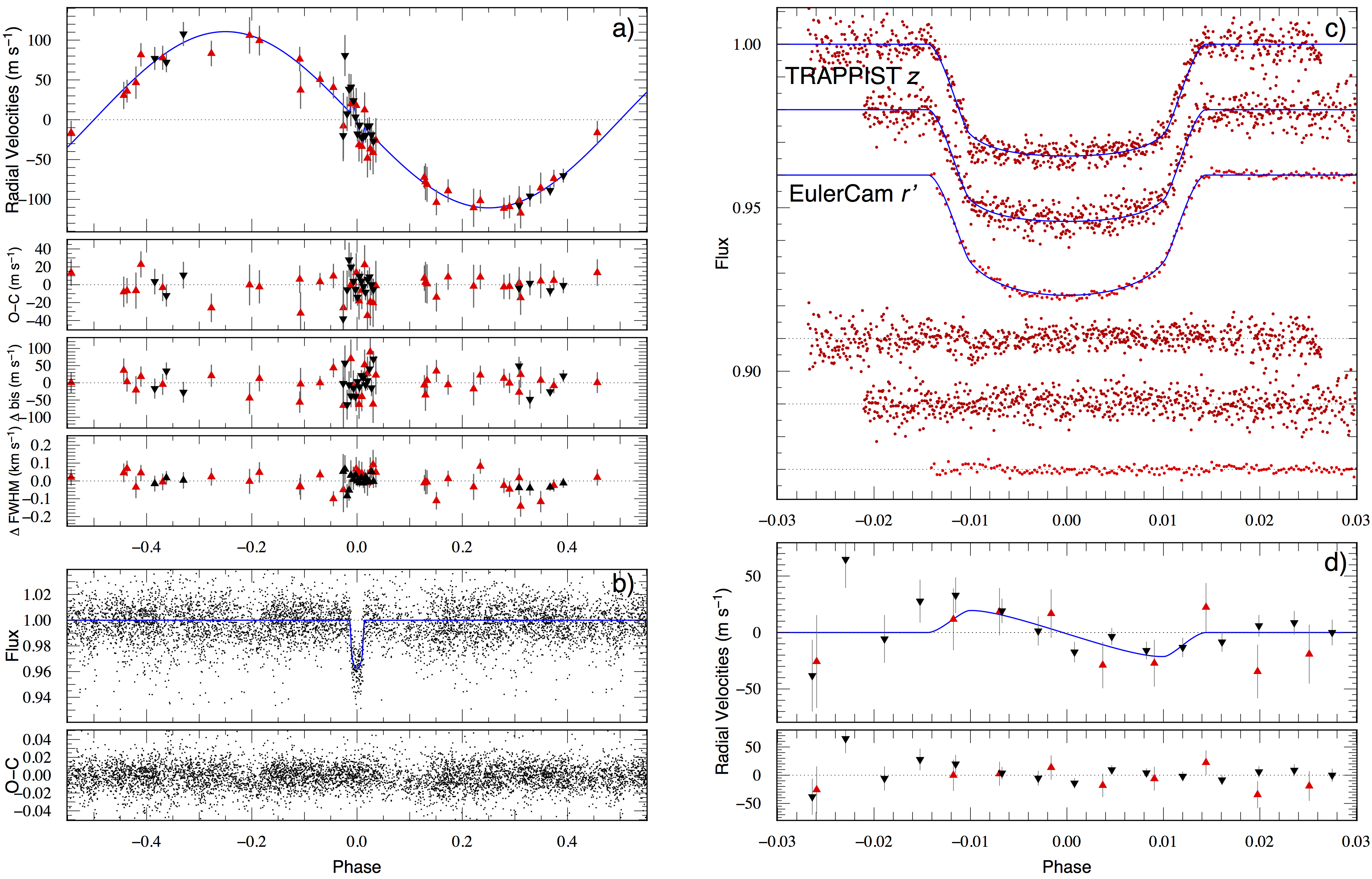}
\caption{a) Top to bottom: CORALIE (upright red triangles) and HARPS (inverted black triangles) radial velocities on \object{WASP-80} plotted with a circular Keplerian model and residuals; 
below: change in the span of the bisector slope and change in the FWHM of the CCF.  b) Phase-folded WASP V+R photometry with model and its residuals. c) Top to bottom: the two TRAPPIST $z$ band and the EulerCam r'-Gunn transit lightcurves with models over plotted. The residuals are displayed in the same
order below. d) Zoom on the Rossiter-McLaughlin effect showing CORALIE  and HARPS radial velocities with the most likely model and the residuals from the fit.}
\label{fig:w80}
\end{figure*}

\object{WASP-80} (\object{1SWASP J201240.26-020838.2}; \object{2MASS J20124017
-0208391}; \object{TYC~5165-481-1}; \object{BPM 80815}) was observed 
5\,782 times during one season at the WASP-South facility in Sutherland (South Africa), 
by a single camera, between 2010 May 15 and 2010 September 26 (Fig. \ref{fig:w80}b). The \textit{Hunter} algorithm \citep{Collier-Cameron:2007pb} found a period 
at 3.07 days and uncovered a transit-like signal from five partial events, with a large depth that nevertheless corresponds to a planet-sized object once the 
colours of the star indicated it was a potential M0 dwarf.

The star was catalogued for spectroscopic follow-up on 2011 May 09, and the first radial velocity measurements were obtained with the CORALIE 
spectrograph in July 2011. Thirty-seven spectra have been collected between 2011 July 21 and 2012 September 12, including ten measurements obtained during the 
Rossiter-McLaughlin effect on 2012 June 19. We also used HARPS and acquired sixteen spectra during the transit of 2012 September 10. Atmospheric conditions 
were poor with seeing $> 2"$ at the beginning of the sequence. Eight measurements were obtained in the nights leading to and following the transit, 
with one point badly affected by weather and excluded from the analysis.
Radial velocities were extracted using a K5 correlation mask and those data also show a 3.07 day variation, in phase with the photometry. 
No such variation can be observed in the span of the bisector slope, or into the width of the line (Fig. \ref{fig:w80}a). This indicates a movement of the 
spectrum with time, as expected for an orbiting planet. 

To complete the 
confirmation of the system and obtain precise physical parameters, three higher precision lightcurves were observed, two by  TRAPPIST \citep{Jehin:2011dk} on 2012 May 07 and 2012 September 10 in the $z$-band, and one by
EulerCam using an r'-Gunn filter on 2012 July 26 (Fig. \ref{fig:w80}c). Our data were reduced and prepared for 
analysis in the same manner as in previous WASP discoveries. Useful references can be found in \citet{Wilson:2008lr}, 
\citet{Anderson:2011fr},  \citet{Gillon:2009qy,Gillon:2012fj}, and \citet{Lendl:2012qy}.








\section{Spectral analysis}\label{sec:spectr}

The HARPS spectra were co-added, leading to a single spectrum of signal-to-noise of 75:1. Its analysis was conducted following the methods described in
\cite{Doyle:2013lr}; the results are displayed in Table \ref{tab:params}. The {\halpha} line being weak,
it indicates a low effective temperature around 4000~K and a K7 spectral type. 
The TiO bands are also weak, typical of a late K--early M dwarf.
The Na {\sc i} D are
very strong, implying a surface gravity (\logg) around 4.6, closer to an M spectral type than a K7 (Fig. \ref{fig:spec}). 
The equivalent widths of several clean and unblended Fe~{\sc i} lines were measured, in order
to determine the stellar metallicity, evaluate microturbulence and confirm the
{\teff} estimated from {\halpha}. 
They were also fitted to estimate the broadening caused by the projected stellar rotation velocity
(\vsini). Macroturbulence was assumed to be zero since its effect is expected to be
lower than thermal broadening \citep{Gray:2008fj}. We found \vsini = $3.55 \pm 0.33$ \kms.
Because this seemed unusually large for this spectral type, we looked at all the stars with a similar $B-V$ present in 
the HARPS archive: \object{WASP-80} has the widest lines in the sample.
There is no significant detection of lithium in the spectrum, and we can place an equivalent-width upper limit of 30m\AA,  meaning log
A(Li) $<$ 0.0 $\pm$ 0.2. For early M-type stars, lithium can be this depleted in less than 100~My \citep{Sestito:2005ys}. 

In their survey of high proper motion stars, \citet{Stephenson:1986qy} listed
the star as spectral type K5. This is, however, inconsistent with the results of
the spectral analysis. An independent check of the stellar temperature can be
obtained from the InfraRed Flux Method (IRFM) \citep{Blackwell:1977uq}, which has been
used to determine {\teff} and stellar angular diameter ($\theta$) by estimating the total observed bolometric flux 
from broad-band photometry from NOMAD, TASS, CMC14, and
2MASS. This gives
\teff\ = 4020 $\pm$ 130~K and $\theta = 0.113 \pm 0.008$~mas, 
consistent with the {\teff} from the spectral analysis. There
is no sign of any interstellar Na~D lines in the spectra, so reddening is
expected to be negligible.   



\begin{table}
\caption{Parameters of the \object{WASP-80} system.}
\begin{tabular}{llll} \hline \hline

\multicolumn{4}{c}{\object{1SWASP J201240.26-020838.2}}\\
\multicolumn{4}{c}{\object{2MASS J20124017-0208391}}\\
\multicolumn{4}{c}{\object{TYC 5165-481-1}}\\
\multicolumn{4}{c}{\object{BPM 80815}}\\

\hline

Filter  & Magnitude  & Filter  &  Magnitude \\ \hline

\multicolumn{2}{l}{\textit{2MASS$^1$}} & \multicolumn{2}{l}{\textit{NOMAD$^2$}}\\
J	&	$9.218 \pm 0.023$	& B	& 12.810\\
H	&	$ 8.513 \pm 0.026$	& V	&11.870\\
K	&	$ 8.351 \pm 0.022$	& R	& 11.110\\
\multicolumn{2}{l}{\textit{TASS4$^3$}} & \multicolumn{2}{l}{\textit{CMC14$^4$}}\\
V	& $11.881\pm 0.228$	& r'	& $11.358 \pm 0.011$ \\
I	& $10.279 \pm 0.105$\\

\hline
\hline

Parameter  & Value \& $1\sigma$ error & Parameter  & Value \& $1\sigma$ error\\ \hline

\multicolumn{4}{l}{\textit{from spectral line analysis}}\\

Spectral type   &   K7V &Distance   &   60 $\pm$ 20 pc \\
\teff      & 4145 $\pm$ 100 K &{[Fe/H]}   &$-$0.14 $\pm$ 0.16 \\
\logg      & 4.6 $\pm$ 0.2 (cgs) &log A(Li)  &   $<$ 0.0 $\pm$ 0.2 \\
\vsini     & 3.55 $\pm$ 0.33 \kms & \mictrb    & 0.3 $\pm$ 0.3 \kms \\
Mass       &   0.58 $\pm$ 0.05 M$_{\odot}$ &Radius     &   0.63 $\pm$ 0.15 R$_{\odot}$ \\
\\
 \multicolumn{4}{l}{\textit{jump parameters for the MCMC}}\\
 Period	&	3.0678504 $^{(+23)}_{(-27)}$ d 	& T$_0$ (BJD)	&	2\,456\,125.417512 $^{(+67)}_{(-52)}$\\
 Depth	&	0.02933 $^{(+10)}_{(-09)}$		& Width	&	0.08800 $^{(+19)}_{(-16)}$ d\\
 $\sqrt{V \sin i_\star} \cos \beta$ 	&	0.48 $^{(+0.12)}_{(-0.13)}$&b		&	0.019 $^{(+26)}_{(-17)}$ R$_\odot$	\\
$\sqrt{V \sin i_\star} \sin \beta$	&	$\pm$1.78 $^{(+0.12)}_{(-0.09)}$& K		&	110.9 $^{(+3.0)}_{(-3.3)}$ m\,s$^{-1}$\\
 \\
 \multicolumn{4}{l}{\textit{derived parameters from the MCMC}}\\
R$_\mathrm{p}$/R$_\star$	&0.17126 $^{(+31)}_{(-26)}$	&f(m)				&	0.425\,10$^{-9}$ $^{(+43)}_{(-31)}$ M$_\odot$\\
R$_\star$/a				&0.07699 $^{(+17)}_{(-17)}	$	&R$_\mathrm{p}$/a		&	0.013183 $^{(+39)}_{(-35)}$\\
$\log\,g_\star$				&4.689 $^{(+12)}_{(-13)}$	 (cgs)&$\log\,g_\mathrm{p}$	&	3.178 $^{(+13)}_{(-12)}$ (cgs)\\
$\rho_\star$				&3.117 $^{(+21)}_{(-20)}$ $\rho_\odot$			&$\rho_\mathrm{p}$	&	0.554 $^{(+30)}_{(-39)}$ $\rho_\mathrm{jup}$\\
M$_\star$	(prior)			&0.57 $^{(+0.05)}_{(-0.05)}$ M$_\odot$	&M$_\mathrm{p}$	&	0.554 $^{(+30)}_{(-39)}$ M$_\mathrm{jup}$\\
R$_\star$					&0.571 $^{(+16)}_{(-16)}$ R$_\odot$	&R$_\mathrm{p}$	&	0.952 $^{(+26)}_{(-27)}$ R$_\mathrm{jup}$\\
$V \sin i_\star$				&3.46 $^{(+0.34)}_{(-0.35)}$	\kms		& $\beta$			&$\pm$75 $^{(+4.0)}_{(-4.3)}$ deg\\
a						&0.0346 $^{(+08)}_{(-11)}$ AU			&$i_\mathrm{p}$	&	89.92 $^{(+0.07)}_{(-0.12)}$ deg\\
e						&$<$ 0.07							&$| \dot\gamma |$	&$< 24$ m\,s$^{-1}$yr$^{-1}$\\
\hline
\end{tabular}
\label{tab:params}
\newline {\bf Note:} Mass and radius estimated using the
\cite{Torres:2010uq} calibration.
Spectral type estimated from \teff\
using Table~B.1 in \citet{Gray:2008fj}.
Distance estimated using the IRFM angular diameter and the stellar radius. Units based on the equatorial solar and jovian radii and masses taken from Allen's Astrophysical Quantities \citep{Cox:2000vn}.  1 - \citet{Skrutskie:2006kx} 2 - \citet{Zacharias:2004fk} 3 - \citet{Droege:2006yq} 4 - ViZier I/304/out

\end{table}

\section{Results}

\begin{figure}
\centering
\includegraphics[width=9cm]{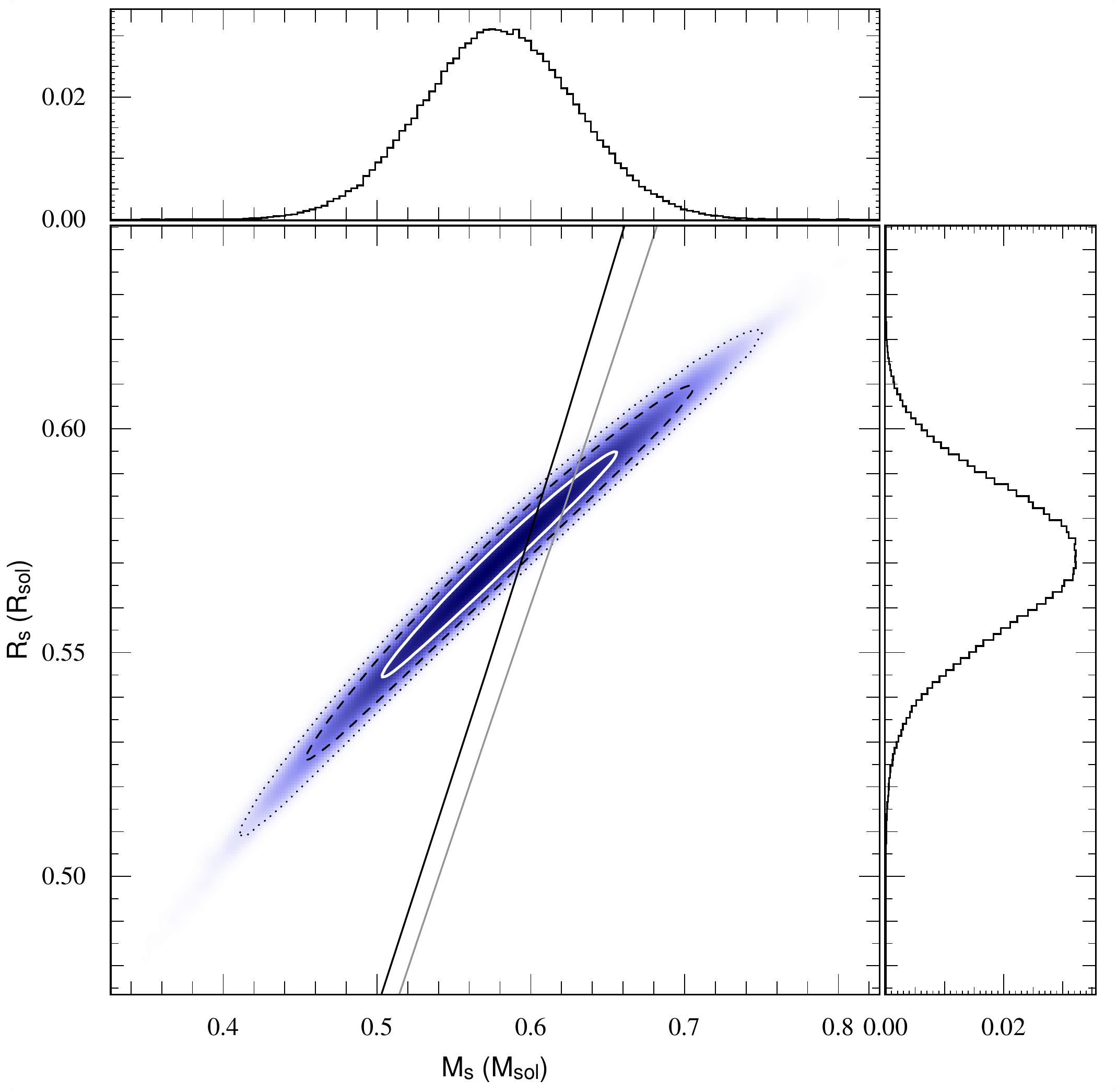}
\caption{Posterior probability density, output of the MCMC, showing the mass of the primary (drawn from a prior), and its radius (inferred from the transit 
signal). 1, 2 and 3 $\sigma$ confidence regions are drawn. The over-plotted grey line is the theoretical 0.5 Gyr mass-radius relationship from 
\citet{Baraffe:1998ly}, in black is the 8 Gyr relation (for solar metallicity). Side histograms show the marginalised parameters.}
\label{fig:W80bar}
\end{figure}

We applied the same fitting MCMC algorithm as described in \citet{Triaud:2011vn}. The stellar mass was constrained by a Gaussian prior with mean 
and standard deviation corresponding to the value obtained using the empirical mass-radius relation from \citet{Torres:2010uq}. The fit of the 
model over the data informs us of the mean stellar density \citep{Seager:2003qy} found to be $3.12\pm0.02\,\rho_\odot$. When combined with a mass of 
$0.58 \pm 0.05 M_\odot$  it gives a radius of $0.57\pm 0.02\, R_\odot$ (entirely compatible with the Torres relation) and a $\log\,g_\star = 4.69\pm 0.02$. 
Those values are also compatible with theoretical mass--radius relationships presented in \citet{Baraffe:1998ly} (see Fig. \ref{fig:W80bar}). 

The fit to the radial-velocities gives a reduced $\chi^2_{\rm r} = 0.75\pm0.17$, suggesting we may have overestimated our error bars on the 
measurements. The largest contribution to the $\chi^2$ comes from the second point in the HARPS Rossiter-McLaughlin series, likely affected by high seeing. 
No eccentricity can be detected; we place a $95\%$ confident upper value at $e< 0.07$. Similarly, if there is any other perturber in the system, it adds an  
acceleration lower than 24 m\,s$^{-1}$yr$^{-1}$. The results of the fit are located in Table \ref{tab:params}. The mass function f(m) and the 
$\log\,g_\mathrm{p}$ are directly obtained from fitting the data; they indicate we have discovered a new transiting planet. Using the stellar 
mass we obtain a mass and radius for our object and find $0.55\pm0.04$ M$_\mathrm{jup}$ and $0.95\pm0.03$ R$_\mathrm{jup}$. 

The Rossiter-McLaughlin effect is marginally detected even though a semi amplitude between 60 and 70 m\,s$^{-1}$ had been expected. This can
be explained either by a highly inclined planet (the impact parameter $b < 0.1$, even when adjusting only the photometry (see \citet{Triaud:2011vn} for details), 
or by an additional unaccounted-for  broadening of the spectral lines. This could occur in the presence of magnetic fields producing a partially
resolved Zeeman line splitting. The presence of strong magnetic
fields has been reported for a number of M dwarfs by \citet{Morin:2010yq} and \citet{Donati:2008kx}. 
We examined our spectra and found no signs of Zeeman broadening.

Two different fits were attempted. For the first one, we assumed $\beta = 0^\circ \pm 20$ and found $V \sin i_\star = 0.91 \pm 0.25$ km\,s$^{-1} $, in strong 
disagreement with the value inferred in section \ref{sec:spectr}. For the second attempt, we chose to impose a prior on $V \sin i_\star = v \sin i_\star$, and 
as expected we find two well separated and symmetrical solutions for the spin--orbit angle: $\beta = \pm 75^\circ  \pm 4$. We caution here that this angle is
entirely dependent on the value of $v \sin i_\star$.

The rotation rate ($P =
8.5 \pm 0.8$~d) implied by the {\vsini} gives a gyrochronological age of $\sim
100^{+30}_{-20}$~My using the \citet{Barnes:2007lr} relation.
The presence of Ca H+K
emission indicates that \object{WASP-80} may be a young active star. Despite this, we do not detect any rotational variability $>$ 1 mmag in the WASP lightcurve or in a multi-epoch TRAPPIST campaign. None of the transits show signs of stellar spot crossings.

Furthermore, using the values of proper motion reported in NOMAD \citep{Zacharias:2004fk} ($-100\pm3$, $-60\pm8$ mas yr$^{-1}$) and the systemic velocity we observed (10.2 km s$^{-1}$), we computed its galactic dynamical velocities (U, V, W)  = (30.0, -11.7, 12.8) km s$^{-1}$. Those values are well away from any of the known young moving groups reported in \citet{Zuckerman:2004qy}. This means the gyrochronological age is not reliable.




\section{Conclusions}

Our observations and their analysis allow us to conclude there is an unseen transiting companion orbiting \object{WASP-80} whose mass and radius are planetary. 


Even though only  a few cool stars have been observed by WASP, our planet confirmation rate is similar to sun-like stars.  We only followed-up 26 stars for which the \textit{Hunter} algorithm returned a signal and whose colours indicate they are of K5 or later type (only three were classified as M0 including \object{WASP-80}). Our data show we have 17 blends (for eight of those the original target was identified as a red giant), three spectrally resolved eclipsing binaries and one unresolved, two potential triples (blends but gravitationally bound), one false alarm, and two planets (WASP-43b \citep{Hellier:2011fk,Gillon:2012fj}, and \object{WASP-80b}). A total of two planets out of 26 candidates is remarkably close to our mean discovery rate of $8.8\pm1.2\%$ planet per candidate \citep{Triaud:2011qy}. 
In addition to the observed rarity of hot Jupiters around cold stars, it is also interesting to note that this planet is orbiting a relatively metal poor star, whereas \citet{Santos:2003lr}, \citet{Fischer:2005qy}, and \citet{Mayor:2011fj} have shown they have an even lower occurrence rate of gas giants for any orbital period. 

\object{WASP-80b}'s equilibrium temperature will be around 800\,K (for an albedo of 0.1). The planet-to-star contrast is favourable for future observation of the emission spectrum of the planet, because it is hosted by a star $\sim$~1500\,K colder than the usual targets. Furthermore, the near $3\%$ depth of the transit makes this gas giant one of the most suitable targets for transmission spectroscopy. 

\object{WASP-80b} is a {\it warm} Jupiter when we consider its temperature, and yet it belongs to the hot Jupiter population. Because of the high density of the host, and the low density of the planet, it is located about 3~Roche radii away from the star, just as would be expected if it had circularised from an earlier, more eccentric orbit \citep{Matsumura:2010ve}. Although the Rossiter-McLaughlin effect is observed symmetrical, the planet's orbital spin could be severely inclined if the host star rotates as quickly as
the spectral line broadening indicates. 
 If no additional spectral line broadening mechanism is discovered, then \object{WASP-80b} would become a rare example of a severely inclined
planet whose host star's $T_{\rm eff}$ is cooler than 6250 K \citep{Brown:2012lr,Albrecht:2012lp,Winn:2010rr}.

\paragraph{\textbf{Nota Bene}}
We used the UTC time standard and Barycentric Julian Dates in our analysis.


\begin{acknowledgements} 
We acknowledge an extensive use of Harvard's ADS paper archive and Strasbourg's CDS data repository. We would like to thank Richard I. Anderson for his kindness in observing the CORALIE Rossiter-McLaughlin effect, with quite late notice and under a certain weather stress, as well as to thank Xavier Bonfils and Corinne Charbonnel for inspiring discussions. This work is supported by the Swiss National Science Foundation.  TRAPPIST is a project funded by the Belgian
Fund for Scientific Research (Fond National de la Recherche Scientifique, F.R.SFNRS)
under grant FRFC 2.5.594.09.F, with the participation of the Swiss
National Science Foundation (SNF). M. Gillon and E. Jehin are FNRS Research
Associates. WASP-South is hosted by the South African
Astronomical Observatory, and we are grateful for their ongoing support
and assistance.

This publication makes use of data products from the Two Micron All Sky Survey, which is a joint project of the University of Massachusetts and the Infrared Processing and Analysis Center/California Institute of Technology, funded by the National Aeronautics and Space Administration and the National Science Foundation.

\end{acknowledgements} 

\bibliographystyle{aa}
\bibliography{../../1Mybib.bib}

\begin{thebibliography}{49}
\expandafter\ifx\csname natexlab\endcsname\relax\def\natexlab#1{#1}\fi

\bibitem[{{Albrecht} {et~al.}(2012){Albrecht}, {Winn}, {Johnson}, {Howard},
  {Marcy}, {Butler}, {Arriagada}, {Crane}, {Shectman}, {Thompson}, {Hirano},
  {Bakos}, \& {Hartman}}]{Albrecht:2012lp}
{Albrecht}, S., {Winn}, J.~N., {Johnson}, J.~A., {et~al.} 2012, \apj, 757, 18

\bibitem[{{Alibert} {et~al.}(2011){Alibert}, {Mordasini}, \&
  {Benz}}]{Alibert:2011fk}
{Alibert}, Y., {Mordasini}, C., \& {Benz}, W. 2011, \aap, 526, A63

\bibitem[{{Anderson} {et~al.}(2011){Anderson}, {Collier Cameron}, {Hellier},
  {Lendl}, {Lister}, {Maxted}, {Queloz}, {Smalley}, {Smith}, {Triaud}, {West},
  {Brown}, {Gillon}, {Pepe}, {Pollacco}, {S{\'e}gransan}, {Street}, \&
  {Udry}}]{Anderson:2011fr}
{Anderson}, D.~R., {Collier Cameron}, A., {Hellier}, C., {et~al.} 2011, \aap,
  531, A60

\bibitem[{{Baraffe} {et~al.}(1998){Baraffe}, {Chabrier}, {Allard}, \&
  {Hauschildt}}]{Baraffe:1998ly}
{Baraffe}, I., {Chabrier}, G., {Allard}, F., \& {Hauschildt}, P.~H. 1998, \aap,
  337, 403

\bibitem[{{Barnes}(2007)}]{Barnes:2007lr}
{Barnes}, S.~A. 2007, \apj, 669, 1167

\bibitem[{{Blackwell} \& {Shallis}(1977)}]{Blackwell:1977uq}
{Blackwell}, D.~E. \& {Shallis}, M.~J. 1977, \mnras, 180, 177

\bibitem[{{Bonfils} {et~al.}(2013){Bonfils}, {Delfosse}, {Udry}, {Forveille},
  {Mayor}, {Perrier}, {Bouchy}, {Gillon}, {Lovis}, {Pepe}, {Queloz}, {Santos},
  {S{\'e}gransan}, \& {Bertaux}}]{Bonfils:2013fk}
{Bonfils}, X., {Delfosse}, X., {Udry}, S., {et~al.} 2013, \aap, 549, A109

\bibitem[{{Brown} {et~al.}(2012){Brown}, {Cameron}, {Anderson}, {Enoch},
  {Hellier}, {Maxted}, {Miller}, {Pollacco}, {Queloz}, {Simpson}, {Smalley},
  {Triaud}, {Boisse}, {Bouchy}, {Gillon}, \& {H{\'e}brard}}]{Brown:2012lr}
{Brown}, D.~J.~A., {Cameron}, A.~C., {Anderson}, D.~R., {et~al.} 2012, \mnras,
  423, 1503

\bibitem[{{Collier Cameron} {et~al.}(2010){Collier Cameron}, {Guenther},
  {Smalley}, {McDonald}, {Hebb}, {Andersen}, {Augusteijn}, {Barros}, {Brown},
  {Cochran}, {Endl}, {Fossey}, {Hartmann}, {Maxted}, {Pollacco}, {Skillen},
  {Telting}, {Waldmann}, \& {West}}]{Collier-Cameron:2010lr}
{Collier Cameron}, A., {Guenther}, E., {Smalley}, B., {et~al.} 2010, \mnras,
  407, 507

\bibitem[{{Collier Cameron} {et~al.}(2007){Collier Cameron}, {Wilson}, {West},
  {Hebb}, {Wang}, {Aigrain}, {Bouchy}, {Christian}, {Clarkson}, {Enoch},
  {Esposito}, {Guenther}, {Haswell}, {H{\'e}brard}, {Hellier}, {Horne},
  {Irwin}, {Kane}, {Loeillet}, {Lister}, {Maxted}, {Mayor}, {Moutou}, {Parley},
  {Pollacco}, {Pont}, {Queloz}, {Ryans}, {Skillen}, {Street}, {Udry}, \&
  {Wheatley}}]{Collier-Cameron:2007pb}
{Collier Cameron}, A., {Wilson}, D.~M., {West}, R.~G., {et~al.} 2007, \mnras,
  380, 1230

\bibitem[{{Cox}(2000)}]{Cox:2000vn}
{Cox}, A.~N. 2000, {Allen's astrophysical quantities}

\bibitem[{{Cumming} {et~al.}(2008){Cumming}, {Butler}, {Marcy}, {Vogt},
  {Wright}, \& {Fischer}}]{Cumming:2008ys}
{Cumming}, A., {Butler}, R.~P., {Marcy}, G.~W., {et~al.} 2008, \pasp, 120, 531

\bibitem[{{Donati} {et~al.}(2008){Donati}, {Morin}, {Petit}, {Delfosse},
  {Forveille}, {Auri{\`e}re}, {Cabanac}, {Dintrans}, {Fares}, {Gastine},
  {Jardine}, {Ligni{\`e}res}, {Paletou}, {Ramirez Velez}, \&
  {Th{\'e}ado}}]{Donati:2008kx}
{Donati}, J.-F., {Morin}, J., {Petit}, P., {et~al.} 2008, \mnras, 390, 545

\bibitem[{{Doyle} {et~al.}(2013){Doyle}, {Smalley}, {Maxted}, {Anderson},
  {Cameron}, {Gillon}, {Hellier}, {Pollacco}, {Queloz}, {Triaud}, \&
  {West}}]{Doyle:2013lr}
{Doyle}, A.~P., {Smalley}, B., {Maxted}, P.~F.~L., {et~al.} 2013, \mnras, 428,
  3164

\bibitem[{{Droege} {et~al.}(2006){Droege}, {Richmond}, {Sallman}, \&
  {Creager}}]{Droege:2006yq}
{Droege}, T.~F., {Richmond}, M.~W., {Sallman}, M.~P., \& {Creager}, R.~P. 2006,
  \pasp, 118, 1666

\bibitem[{{Fischer} \& {Valenti}(2005)}]{Fischer:2005qy}
{Fischer}, D.~A. \& {Valenti}, J. 2005, \apj, 622, 1102

\bibitem[{{Gillon} {et~al.}(2009){Gillon}, {Anderson}, {Triaud}, {Hellier},
  {Maxted}, {Pollaco}, {Queloz}, {Smalley}, {West}, {Wilson}, {Bentley},
  {Collier Cameron}, {Enoch}, {Hebb}, {Horne}, {Irwin}, {Joshi}, {Lister},
  {Mayor}, {Pepe}, {Parley}, {Segransan}, {Udry}, \&
  {Wheatley}}]{Gillon:2009qy}
{Gillon}, M., {Anderson}, D.~R., {Triaud}, A.~H.~M.~J., {et~al.} 2009, \aap,
  501, 785

\bibitem[{{Gillon} {et~al.}(2012){Gillon}, {Triaud}, {Fortney}, {Demory},
  {Jehin}, {Lendl}, {Magain}, {Kabath}, {Queloz}, {Alonso}, {Anderson},
  {Collier Cameron}, {Fumel}, {Hebb}, {Hellier}, {Lanotte}, {Maxted},
  {Mowlavi}, \& {Smalley}}]{Gillon:2012fj}
{Gillon}, M., {Triaud}, A.~H.~M.~J., {Fortney}, J.~J., {et~al.} 2012, \aap,
  542, A4

\bibitem[{{Gray}(2008)}]{Gray:2008fj}
{Gray}, D.~F. 2008, {The Observation and Analysis of Stellar Photospheres}, ed.
  {Gray, D.~F.}

\bibitem[{{Hebb} {et~al.}(2009){Hebb}, {Collier-Cameron}, {Loeillet},
  {Pollacco}, {H{\'e}brard}, {Street}, {Bouchy}, {Stempels}, {Moutou},
  {Simpson}, {Udry}, {Joshi}, {West}, {Skillen}, {Wilson}, {McDonald},
  {Gibson}, {Aigrain}, {Anderson}, {Benn}, {Christian}, {Enoch}, {Haswell},
  {Hellier}, {Horne}, {Irwin}, {Lister}, {Maxted}, {Mayor}, {Norton}, {Parley},
  {Pont}, {Queloz}, {Smalley}, \& {Wheatley}}]{Hebb:2009lr}
{Hebb}, L., {Collier-Cameron}, A., {Loeillet}, B., {et~al.} 2009, \apj, 693,
  1920

\bibitem[{{Hebb} {et~al.}(2010){Hebb}, {Collier-Cameron}, {Triaud}, {Lister},
  {Smalley}, {Maxted}, {Hellier}, {Anderson}, {Pollacco}, {Gillon}, {Queloz},
  {West}, {Bentley}, {Enoch}, {Haswell}, {Horne}, {Mayor}, {Pepe}, {Segransan},
  {Skillen}, {Udry}, \& {Wheatley}}]{Hebb:2010fk}
{Hebb}, L., {Collier-Cameron}, A., {Triaud}, A.~H.~M.~J., {et~al.} 2010, \apj,
  708, 224

\bibitem[{{Hellier} {et~al.}(2009){Hellier}, {Anderson}, {Collier Cameron},
  {Gillon}, {Hebb}, {Maxted}, {Queloz}, {Smalley}, {Triaud}, {West}, {Wilson},
  {Bentley}, {Enoch}, {Horne}, {Irwin}, {Lister}, {Mayor}, {Parley}, {Pepe},
  {Pollacco}, {Segransan}, {Udry}, \& {Wheatley}}]{Hellier:2009lr}
{Hellier}, C., {Anderson}, D.~R., {Collier Cameron}, A., {et~al.} 2009, \nat,
  460, 1098

\bibitem[{{Hellier} {et~al.}(2011){Hellier}, {Anderson}, {Collier Cameron},
  {Gillon}, {Jehin}, {Lendl}, {Maxted}, {Pepe}, {Pollacco}, {Queloz},
  {S{\'e}gransan}, {Smalley}, {Smith}, {Southworth}, {Triaud}, {Udry}, \&
  {West}}]{Hellier:2011fk}
{Hellier}, C., {Anderson}, D.~R., {Collier Cameron}, A., {et~al.} 2011, \aap,
  535, L7

\bibitem[{{Howard} {et~al.}(2012){Howard}, {Marcy}, {Bryson}, {Jenkins},
  {Rowe}, {Batalha}, {Borucki}, {Koch}, {Dunham}, {Gautier}, {Van Cleve},
  {Cochran}, {Latham}, {Lissauer}, {Torres}, {Brown}, {Gilliland}, {Buchhave},
  {Caldwell}, {Christensen-Dalsgaard}, {Ciardi}, {Fressin}, {Haas}, {Howell},
  {Kjeldsen}, {Seager}, {Rogers}, {Sasselov}, {Steffen}, {Basri},
  {Charbonneau}, {Christiansen}, {Clarke}, {Dupree}, {Fabrycky}, {Fischer},
  {Ford}, {Fortney}, {Tarter}, {Girouard}, {Holman}, {Johnson}, {Klaus},
  {Machalek}, {Moorhead}, {Morehead}, {Ragozzine}, {Tenenbaum}, {Twicken},
  {Quinn}, {Isaacson}, {Shporer}, {Lucas}, {Walkowicz}, {Welsh}, {Boss},
  {Devore}, {Gould}, {Smith}, {Morris}, {Prsa}, {Morton}, {Still}, {Thompson},
  {Mullally}, {Endl}, \& {MacQueen}}]{Howard:2012ul}
{Howard}, A.~W., {Marcy}, G.~W., {Bryson}, S.~T., {et~al.} 2012, \apjs, 201, 15

\bibitem[{{Ida} \& {Lin}(2005)}]{Ida:2005lr}
{Ida}, S. \& {Lin}, D.~N.~C. 2005, \apj, 626, 1045

\bibitem[{{Jehin} {et~al.}(2011){Jehin}, {Gillon}, {Queloz}, {Magain},
  {Manfroid}, {Chantry}, {Lendl}, {Hutsem{\'e}kers}, \& {Udry}}]{Jehin:2011dk}
{Jehin}, E., {Gillon}, M., {Queloz}, D., {et~al.} 2011, The Messenger, 145, 2

\bibitem[{{Johnson} {et~al.}(2012){Johnson}, {Gazak}, {Apps}, {Muirhead},
  {Crepp}, {Crossfield}, {Boyajian}, {von Braun}, {Rojas-Ayala}, {Howard},
  {Covey}, {Schlawin}, {Hamren}, {Morton}, {Marcy}, \&
  {Lloyd}}]{Johnson:2012vn}
{Johnson}, J.~A., {Gazak}, J.~Z., {Apps}, K., {et~al.} 2012, \aj, 143, 111

\bibitem[{{Johnson} {et~al.}(2010){Johnson}, {Howard}, {Marcy}, {Bowler},
  {Henry}, {Fischer}, {Apps}, {Isaacson}, \& {Wright}}]{Johnson:2010ly}
{Johnson}, J.~A., {Howard}, A.~W., {Marcy}, G.~W., {et~al.} 2010, \pasp, 122,
  149

\bibitem[{{Laughlin} {et~al.}(2004){Laughlin}, {Bodenheimer}, \&
  {Adams}}]{Laughlin:2004gf}
{Laughlin}, G., {Bodenheimer}, P., \& {Adams}, F.~C. 2004, \apjl, 612, L73

\bibitem[{{Lendl} {et~al.}(2012){Lendl}, {Anderson}, {Collier-Cameron},
  {Doyle}, {Gillon}, {Hellier}, {Jehin}, {Lister}, {Maxted}, {Pepe},
  {Pollacco}, {Queloz}, {Smalley}, {S{\'e}gransan}, {Smith}, {Triaud}, {Udry},
  {West}, \& {Wheatley}}]{Lendl:2012qy}
{Lendl}, M., {Anderson}, D.~R., {Collier-Cameron}, A., {et~al.} 2012, \aap,
  544, A72

\bibitem[{{Matsumura} {et~al.}(2010){Matsumura}, {Peale}, \&
  {Rasio}}]{Matsumura:2010ve}
{Matsumura}, S., {Peale}, S.~J., \& {Rasio}, F.~A. 2010, \apj, 725, 1995

\bibitem[{{Mayor} {et~al.}(2011){Mayor}, {Marmier}, {Lovis}, {Udry},
  {S{\'e}gransan}, {Pepe}, {Benz}, {Bertaux}, {Bouchy}, {Dumusque}, {Lo Curto},
  {Mordasini}, {Queloz}, \& {Santos}}]{Mayor:2011fj}
{Mayor}, M., {Marmier}, M., {Lovis}, C., {et~al.} 2011, eprint arXiv:1109.2497

\bibitem[{{Mayor} \& {Queloz}(1995)}]{Mayor:1995uq}
{Mayor}, M. \& {Queloz}, D. 1995, \nat, 378, 355

\bibitem[{{Mordasini} {et~al.}(2012){Mordasini}, {Alibert}, {Benz}, {Klahr}, \&
  {Henning}}]{Mordasini:2012ve}
{Mordasini}, C., {Alibert}, Y., {Benz}, W., {Klahr}, H., \& {Henning}, T. 2012,
  \aap, 541, A97

\bibitem[{{Morin} {et~al.}(2010){Morin}, {Donati}, {Petit}, {Delfosse},
  {Forveille}, \& {Jardine}}]{Morin:2010yq}
{Morin}, J., {Donati}, J.-F., {Petit}, P., {et~al.} 2010, \mnras, 407, 2269

\bibitem[{{Nutzman} \& {Charbonneau}(2008)}]{Nutzman:2008qy}
{Nutzman}, P. \& {Charbonneau}, D. 2008, \pasp, 120, 317

\bibitem[{{Pollacco} {et~al.}(2006){Pollacco}, {Skillen}, {Collier Cameron},
  {Christian}, {Hellier}, {Irwin}, {Lister}, {Street}, {West}, {Anderson},
  {Clarkson}, {Deeg}, {Enoch}, {Evans}, {Fitzsimmons}, {Haswell}, {Hodgkin},
  {Horne}, {Kane}, {Keenan}, {Maxted}, {Norton}, {Osborne}, {Parley}, {Ryans},
  {Smalley}, {Wheatley}, \& {Wilson}}]{Pollacco:2006fj}
{Pollacco}, D.~L., {Skillen}, I., {Collier Cameron}, A., {et~al.} 2006, \pasp,
  118, 1407

\bibitem[{{Santos} {et~al.}(2003){Santos}, {Israelian}, {Mayor}, {Rebolo}, \&
  {Udry}}]{Santos:2003lr}
{Santos}, N.~C., {Israelian}, G., {Mayor}, M., {Rebolo}, R., \& {Udry}, S.
  2003, \aap, 398, 363

\bibitem[{{Seager} \& {Mall{\'e}n-Ornelas}(2003)}]{Seager:2003qy}
{Seager}, S. \& {Mall{\'e}n-Ornelas}, G. 2003, \apj, 585, 1038

\bibitem[{{Sestito} \& {Randich}(2005)}]{Sestito:2005ys}
{Sestito}, P. \& {Randich}, S. 2005, \aap, 442, 615

\bibitem[{{Skrutskie} {et~al.}(2006){Skrutskie}, {Cutri}, {Stiening},
  {Weinberg}, {Schneider}, {Carpenter}, {Beichman}, {Capps}, {Chester},
  {Elias}, {Huchra}, {Liebert}, {Lonsdale}, {Monet}, {Price}, {Seitzer},
  {Jarrett}, {Kirkpatrick}, {Gizis}, {Howard}, {Evans}, {Fowler}, {Fullmer},
  {Hurt}, {Light}, {Kopan}, {Marsh}, {McCallon}, {Tam}, {Van Dyk}, \&
  {Wheelock}}]{Skrutskie:2006kx}
{Skrutskie}, M.~F., {Cutri}, R.~M., {Stiening}, R., {et~al.} 2006, \aj, 131,
  1163

\bibitem[{{Stephenson}(1986)}]{Stephenson:1986qy}
{Stephenson}, C.~B. 1986, \aj, 92, 139

\bibitem[{{Torres} {et~al.}(2010){Torres}, {Andersen}, \&
  {Gim{\'e}nez}}]{Torres:2010uq}
{Torres}, G., {Andersen}, J., \& {Gim{\'e}nez}, A. 2010, \aapr, 18, 67

\bibitem[{{Triaud}(2011)}]{Triaud:2011qy}
{Triaud}, A.~H.~M.~J. 2011, PhD thesis, Observatoire Astronomique de
  l'Universite de Geneve, http://archive-ouverte.unige.ch/unige:18065

\bibitem[{{Triaud} {et~al.}(2011){Triaud}, {Queloz}, {Hellier}, {Gillon},
  {Smalley}, {Hebb}, {Collier Cameron}, {Anderson}, {Boisse}, {H{\'e}brard},
  {Jehin}, {Lister}, {Lovis}, {Maxted}, {Pepe}, {Pollacco}, {S{\'e}gransan},
  {Simpson}, {Udry}, \& {West}}]{Triaud:2011vn}
{Triaud}, A.~H.~M.~J., {Queloz}, D., {Hellier}, C., {et~al.} 2011, \aap, 531,
  A24

\bibitem[{{Wilson} {et~al.}(2008){Wilson}, {Gillon}, {Hellier}, {Maxted},
  {Pepe}, {Queloz}, {Anderson}, {Collier Cameron}, {Smalley}, {Lister},
  {Bentley}, {Blecha}, {Christian}, {Enoch}, {Haswell}, {Hebb}, {Horne},
  {Irwin}, {Joshi}, {Kane}, {Marmier}, {Mayor}, {Parley}, {Pollacco}, {Pont},
  {Ryans}, {Segransan}, {Skillen}, {Street}, {Udry}, {West}, \&
  {Wheatley}}]{Wilson:2008lr}
{Wilson}, D.~M., {Gillon}, M., {Hellier}, C., {et~al.} 2008, \apjl, 675, L113

\bibitem[{{Winn} {et~al.}(2010){Winn}, {Fabrycky}, {Albrecht}, \&
  {Johnson}}]{Winn:2010rr}
{Winn}, J.~N., {Fabrycky}, D., {Albrecht}, S., \& {Johnson}, J.~A. 2010, \apjl,
  718, L145

\bibitem[{{Zacharias} {et~al.}(2004){Zacharias}, {Monet}, {Levine}, {Urban},
  {Gaume}, \& {Wycoff}}]{Zacharias:2004fk}
{Zacharias}, N., {Monet}, D.~G., {Levine}, S.~E., {et~al.} 2004, in Bulletin of
  the American Astronomical Society, Vol.~36, American Astronomical Society
  Meeting Abstracts, 1418

\bibitem[{{Zuckerman} \& {Song}(2004)}]{Zuckerman:2004qy}
{Zuckerman}, B. \& {Song}, I. 2004, \araa, 42, 685

\end{thebibliography}
\end{document}